\documentclass[runningheads,openany]{svmult}
\usepackage[T1]{fontenc}
\usepackage{palatino}
\usepackage{euler}
\usepackage{graphicx}  
\usepackage{physprbb}  
\usepackage{bg2019}  

\begin{document}
\title*{The enigmatic $\Delta(1600)$ resonance}


\author{B. Golli}
\institute{
Faculty of Education,
              University of Ljubljana
 and
Jo\v{z}ef Stefan Institute, 
              1000 Ljubljana, Slovenia}

\maketitle

\begin{abstract}
  Our recently proposed model of the $\Delta(1600)$ resonance,
  in which the dominant component is a quasi-bound state of the
  $\Delta(1232)$ and the pion, is confronted with
  a similar model of the $N^*(1440)$ resonance
  as its counterpart in the P11 partial wave.
  We stress an essentially different mechanism
  responsible for generating the two resonances.
\end{abstract}

\noindent
The two low-lying resonances in the P11 and P33 partial waves,
the Roper resonance ($N^*(1440)$) and the $\Delta(1600)$ resonance,
have been attracting special attention due to their relatively low
masses compared to the prediction of the quark model in which
they figure as the first radial excitations in the respective
channel, and have been considered
as candidates for dynamically generated resonances.
In order to understand the mechanism of their formation
we study these two resonances in
a chiral quark model,
which may produce either a genuine resonance by exciting the quark
core, or a dynamically generated resonance involving a baryon-meson
quasi-bound state.
We use a coupled channel approach involving the $\pi N$, $\pi\Delta$,
and $\sigma N$ channels which --- based on our previous experience ---
dominate the intermediate energy regime in the P11 and P33 partial
waves.
The Cloudy Bag Model (CBM) is used to fix the quark-pion vertices while 
the $s$-wave $\sigma$-baryon vertex is introduced phenomenologically 
with the coupling constant $g_\sigma$ as a free parameter.
Labeling the channels by $\alpha,\beta,\gamma$,
the Lippmann-Schwinger equation for the meson amplitude $\chi_{\alpha\gamma}$
for the process $\gamma\to\alpha$ can be cast in the form
\begin{equation}
  {\chi}_{\alpha\gamma}(k_\alpha,k_\gamma) 
         = {\mathcal{K}_{\alpha\gamma}}(k_\alpha,k_\gamma)
+ \sum_\beta\int{\rm d} k\;
  {{\mathcal{K}_{\alpha\beta}}(k_\alpha,k){\chi}_{\beta\gamma}(k,k_\gamma)
  \over \omega(k) + E_{\beta}(k)-W}\,.
\label{eq4chi}
\end{equation}
The half-on-shell pion amplitude consists of the resonant and
non-resonant part,
\begin{equation}
   \chi_{\alpha\gamma}(k,k_\gamma) = 
    c_{\gamma R}{\cal V}_{\alpha R}(k)+
    {\cal D}_{\alpha\gamma}(k,k_\gamma)\,,
\label{splitchi}
\end{equation}
with the non-resonant part ${\cal D}_{\alpha\gamma}(k,k_\gamma)$ satisfying 
the same Lippmann-Schwinger equation, while the dressed vertex 
${\cal V}_{\alpha R}(k)$ satisties the Lippmann-Schwinger equation with 
the same kernel and the bare vertex for the non-homogeneous part.
Approximating the kernel $\mathcal{K}$ by a separable form,
the integral equations reduce to a system of linear equations
which can be solved exactly.
The resulting amplitude is proportional to the $K$ matrix which,
in turn, determines the scattering $T$ matrix.
The Laurent-Pietarinen expansion is finally used to extract the 
information about the $S$-matrix poles in the complex energy plane.

\begin{figure}[h]
\begin{minipage}[t]{46mm}
\caption{\label{fig:polesR}
Evolution of the $N^*(1440)$ mass (Re$\,W$) and the width
(proportional to the radius of the circle) as a function of 
the interaction strength for two bare masses of
the three-quark configuration, 1750~MeV and 2000~MeV;
$g/g_0$ denotes the reduction factor, 
equal for each coupling constant.
The radius at $g/g_0=1$ corresponds to Im$\,W=180$~MeV.}
\end{minipage}
\hfill
\begin{minipage}[t]{73mm}
\hrule height 0pt depth 0pt
\vspace{6pt}
\includegraphics[width=72mm]{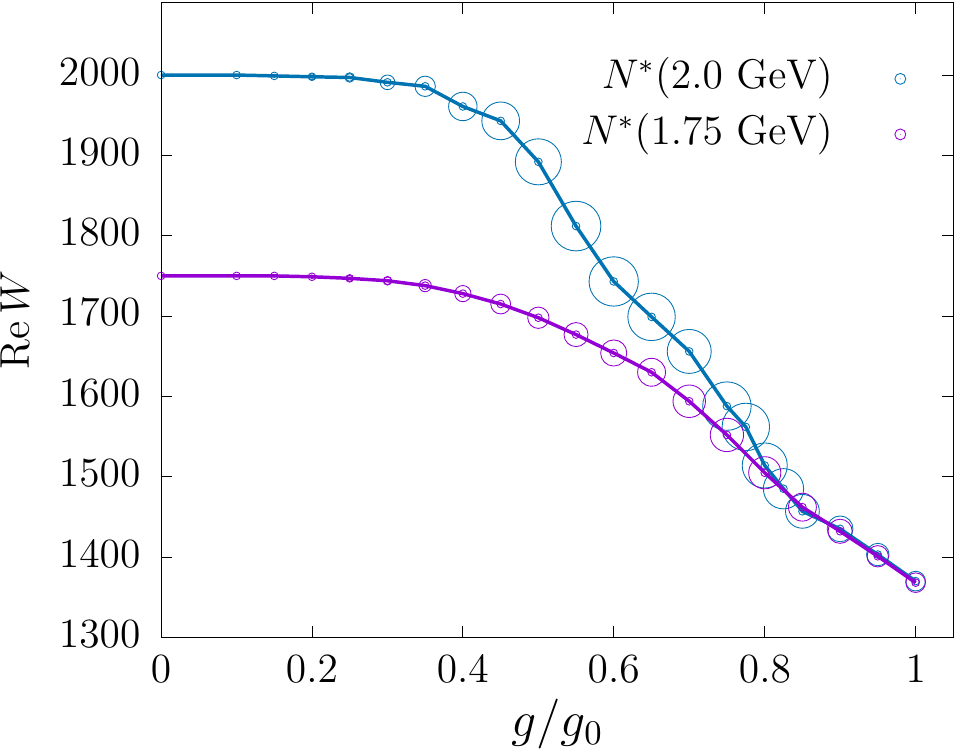}
\end{minipage}
\end{figure}

The formation of the Roper resonance ($N^*(1440)$) is studied in
Ref.~\cite{PRC18}, confronting two mechanisms for resonance
formation: the explicit inclusion of a resonant three-quark state
in which one quark is promoted to the $2s$ state, and the
dynamical generation in the absence of the resonant state.
In both cases the nucleon pole is explicitly included.
While the $p$-wave $\pi N$ interaction is repulsive in the P11 
channel, the $s$-wave $\sigma N$ interaction is attractive, 
and is able to support a (quasi) bound state for sufficiently 
strong $g_\sigma$.
The resulting mass of the resonance is close to the PDG
value in a relatively wide interval of  $g_\sigma$, while its width
is smaller than the PDG value and drops with increasing  $g_\sigma$.
Including a three-quark resonant state, the mass of the resonance 
remains almost the same, while its width increases and comes very 
close to its PDG value (see Table III in \cite{PRC18}).
The result is rather insensitive to the mass of the three-quark
resonant state, which allows us to use a value around 2~GeV,
in agreement with the quark-model ordering of the $2s$ and 
$1p$ states, as well as with the recent results of the lattice 
calculations \cite{lang16,kiratidis17} which have not found 
a sizable three-quark component  below $\sim\!\!1.7$~GeV.
We conclude that while the mass of the $S$-matrix pole is 
determined by the dynamically generated state,
its width and modulus are strongly influenced by the three-quark
resonant state.
This conclusion is further supported by a smooth evolution
of the $S$-matrix pole in the complex energy plane as the coupling
of the $\sigma$ as well as of the pion to the quark core is gradually 
increased on (see Fig.~\ref{fig:polesR}).
Starting with two bare masses of 1750~MeV and 2000~MeV,
both curves end up almost at the same point
with the mass and width consistent with the PDG values.

Though we might expect that, because of apparently the same
three-quark configuration, the situation with the $\Delta(1600)$ is 
similar to that with the $N^*(1440)$ resonance, this is not the case.
One important difference is the nature of the $p$-wave $\pi N$
interaction which is attractive in the P33 partial wave, in contrast
to its repulsive character in the P11, P13, and P31 waves.
Furthermore, the analog of the $\sigma N$ system, the 
$\sigma\Delta(1232)$ system, turns out to make a sizable 
contribution to the scattering amplitude only above 1700~MeV, 
and hence the $\sigma$ plays a minor role in the formation of 
the $\Delta(1600)$ resonance.
In~\cite{PRC19} we therefore consider only the $\pi N$ and
the $\pi\Delta$ channels.
Since the $\pi N$ 
coupling constant
is fixed by the behavior of the scattering amplitudes near the 
threshold, the only free parameter in the underlying model (CBM) 
is the bag radius $R$ which is inversely proportional to the 
cutoff energy; for the value of $R=0.8$~fm, leading to the most 
consistent results for the nucleon as well as for the low lying 
resonances, it corresponds to $\approx 550$~MeV.

\begin{figure}[h]
\begin{minipage}[t]{40mm}
\caption{\label{fig:polesDl}
Evolution of the poles as a function of the bag radius
in the P33 partial wave in three different approximations:
(i) including only the nucleon and the pion (orange
curve and circles), 
(ii) including the nucleon and the $\Delta$ but without 
a resonant state (green), 
(iii) with the $\Delta$ resonant states (red).
The width of the resonance $-2{\rm Im}\,W$ is proportional
to the radius of the circle.}
\end{minipage}
\hfill
\begin{minipage}[t]{78mm}
\hrule height 0pt depth 0pt
\vspace{6pt}
\includegraphics[width=80mm]{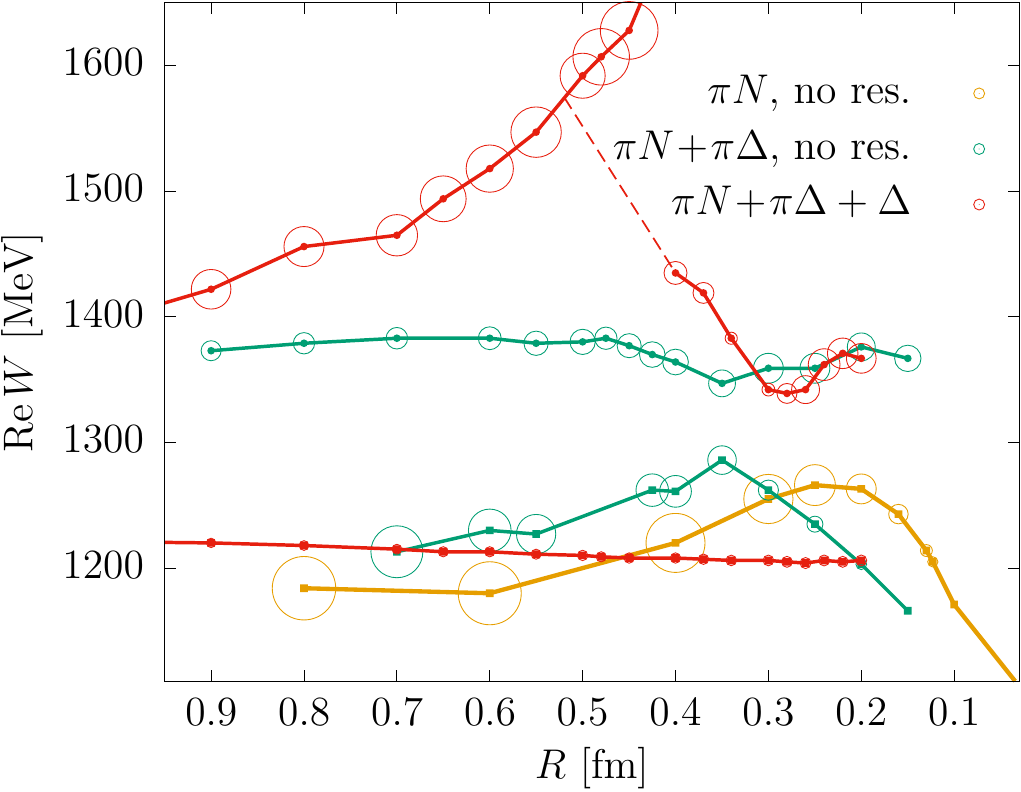}
\end{minipage}
\end{figure}

Already a few years after the discovery of the $\Delta(1232)$
resonance, it was conjectured that this resonance arises as a
consequence of the attraction in the $\pi N$ system at sufficiently
strong cutoff~\cite{ChewLow56}.
In our model we do observe a resonance in the $\pi N$ system
manifesting itself as a pole in the complex energy plane at a mass
around 1200~MeV, with a width that decreases with increasing interaction
strength (decreasing $R$) (orange curve in Fig.~\ref{fig:polesDl}).
For $R=0.123$~fm the mass and the width reach the values which agree 
well with the PDG values, and for $R=0.050$~fm the system becomes bound.
We next include the $\Delta$ (in addition to the nucleon) as
the $u$-channel exchange particle in the kernel, and solve 
(\ref{eq4chi}) for the nonresonant amplitude $\mathcal{D}$.
Besides the pole at around 1200~MeV another pole slightly
below 1400~MeV emerges (green curves in  Fig.~\ref{fig:polesDl}).
The second pole is dominated by the $\pi\Delta$ configuration and
can be interpreted as a progenitor of the $\Delta(1600)$ resonance.

We next include a three-quark state corresponding to the $\Delta(1232)$
in the $s$-channel and fix its bare mass such that the resulting
Breit-Wigner mass (i.e., the zero of Re$\,T$) appears at  1232~MeV.
With decreasing $R$ the resonant state mixes more and more strongly
with the lower dynamically generated state, forming the physical
$\Delta(1232)$.
The latter component dominates below $R=0.2$~fm, nonetheless, the mass 
and the width of the resonance pole remain constant (red curves 
in Fig.~\ref{fig:polesDl}) and stay close to the PDG value.
The upper dynamically generated resonance is pushed toward
a slightly higher mass and acquires a larger width.
In the physically sensible region around $R\approx0.8$~fm,
the mass and the width come close to the PDG values for the 
$\Delta(1600)$ resonance.
The attribution of this pole to the $\Delta(1600)$ resonance
is, however, not justified for smaller $R$, where its mass
keeps increasing, and, in addition, another branch emerges, 
approaching the upper dynamically generated resonance.

We finally add a bare $(1s)^2(2s)$ configuration representing
the first radial excitation of the $\Delta(1232)$.
In the harmonic oscillator model, its mass is expected to
lie $\sim\!\!1$~GeV above the $(1s)^3$ configuration,
so we fix its (bare) mass at 2.2~GeV,
while its coupling is taken from the CBM.
Apart from the two resonances discussed above, the third resonance
emerges with a mass (Re$\,W$) close to the bare value.
Increasing the strength of the interaction 
(decreasing $R$) we notice that it
stays almost constant and --- at least in the physically
relevant regime of $R$'s --- well separated from the
other two resonances.
\begin{figure}[h]
\begin{minipage}[t]{38mm}
\caption{\label{fig:polesDu}
Evolution of the poles in the model including two resonant states
with the second state at the bare mass of 2.2~GeV (blue curves)
and at 2.0~GeV (violet), respectively, compared with the model
involving  $\Delta$ alone (red, the same curve as 
in Fig.~\ref{fig:polesDl}).
}
\end{minipage}
\hfill
\begin{minipage}[t]{78mm}
\hrule height 0pt depth 0pt
\vspace{6pt}
\includegraphics[width=75mm]{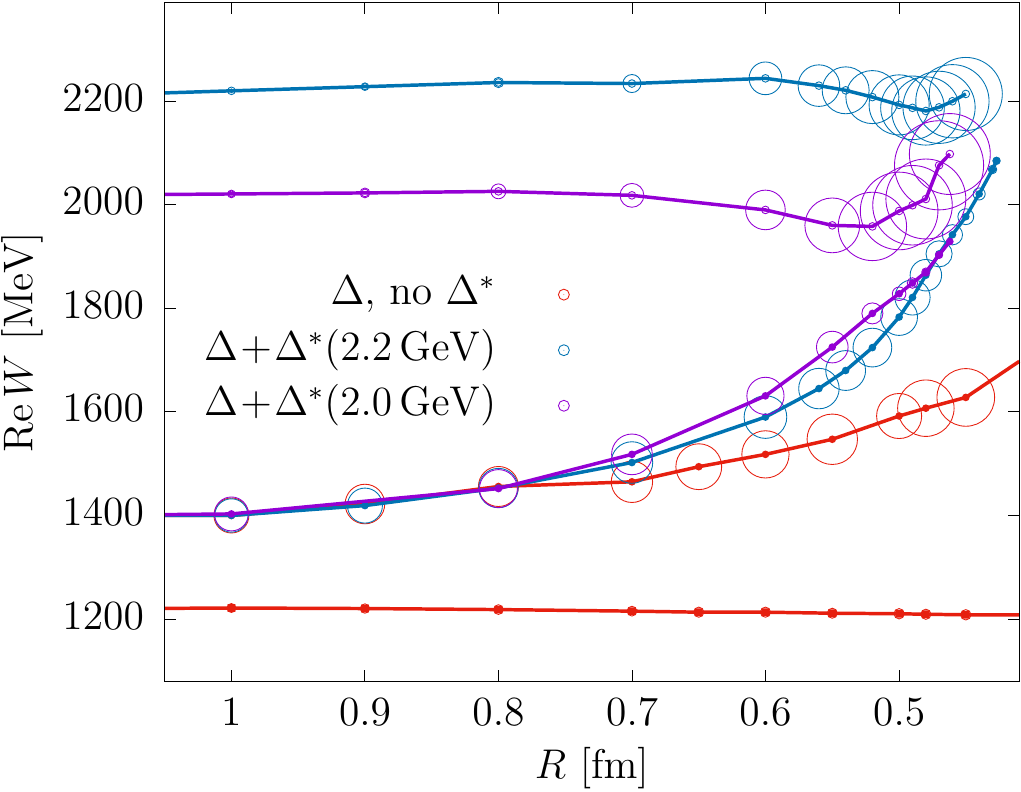}
\end{minipage}
\end{figure}

We can therefore conclude that the radially excited quark state plays 
a very minor role in the formation of the $\Delta(1600)$ resonance,
which in our model turns out to be primarily a quasi-bound
state of $\Delta(1232)$ and the pion.
This mechanism is therefore fundamentally different from that
responsible for the formation of the $N^*(1440)$ resonance,
discussed above, and originates in the different nature of
the pion interaction in the two partial waves.

This work has been done in collaboration with H. Osmanovi\'c (Tuzla)
and S. \v{S}irca (Ljubljana).

\end{document}